# Graph Based Power Flow Calculation for Energy Management System

Junjie Shi[a], Guangyi Liu[b], *Senior Member, IEEE*, Renchang Dai[b], *Senior Member, IEEE*
Jingjin Wu[b], Chen Yuan[b], Zhiwei Wang[b]
[a] State Grid Sichun Electric Power Company, Sichuan, China
[b] Global Energy Interconnection Research Institute North America, San Jose, CA, USA
renchang.dai@geirina.net

*Abstract*—Power flow calculation in EMS is required to accommodate large and complex power system. To achieve a faster than real-time calculation, a graph based power flow calculation is proposed in this paper. Graph database and graph computing advantages in power system calculations are presented. A linear solver for power flow application is formulated and decomposed in nodal parallelism and hierarchical parallelism to fully utilize graph parallel computing capability. Comparison of the algorithm with traditional sequential programs shows significant benefits on computation efficiency. Case studies on practical large-scale systems provide supporting evidence that the new algorithm is promising for online computing for EMS.

## I. Introduction

Electrical power system is revolutionizing over decades into a highly interconnected, large and complex network. Populations and economic growth globally demands more electricity. Transactions crossing large areas are encouraged to make more economic and environmental sense, and result in large power flowing over a wide area. High voltage transmission technologies boosted voltage level to 1000kV for Ultra High Voltage Alternating Current (UHVAC) and ±1000kV Ultra High Voltage Direct Current (UHVDC) to transmit power over thousand miles [1]. Advanced power electronics devices enable Flexible Alternating Current Transmission Systems (FACTS), for instance Static Var Compensator and voltage source converter based STATCOM, being adapted to control power flow agilely and accurately in electric power grid [2-3]. The economic growth and technology development encourage the ambitions of building globally interconnected energy network [4].

High penetration intermittent renewable energy resources on the generation side are interconnected to both transmission network and distribution network interfacing with power electronics devices[5-8]. As more renewable generation (e.g. wind turbines, solar photovoltaic) and demand response penetrated, system synchronous inertia declining is observed worldwide [9] requiring a fast response on frequency regulation. Varied storage facilities and demand response are widely adopted to benefit variable energy resources and uncertain load demands [10-14] meanwhile bringing higher complexity. The adaptation of UHVAC, UHVDC, FACTS, renewable energy, demand response, and storage facility dramatically increases the complexity of nowadays electric power system.

The trends of the electric power system are challenging existing power flow application in Energy Management System (EMS). Power flow calculation is required to be evolving to accommodate larger scale, higher complex, more constrained and uncertain power system with a faster than real-time manner or even look-ahead capability with future situational awareness [15].

Power system society has endeavored to redesign power flow application from database structure to advanced algorithms to speed up power flow calculation for large-scale power systems. To achieve a faster than real-time analysis, a novel system architecture and fast computational method are needed to assist operators to ensure a reliable, resilient, secure and efficient electric power grid in a timely manner. Among the various computational procedures, parallel computing is a promising technology to improve computation efficiency taking advantages of modern computation technology, abundant storage space and parallel capability of database and processing units. However, the state of art of power flow application does not effectively harness the parallel capability for the reason that the traditional relational database and computation methods applied by power flow application in EMS were not designed for parallel computing.

To accommodate parallel computing, database and calculation method for power flow application need to be redesigned to fit into parallel database management, parallel analysis, and fast visualization.

When performing power flow calculation, from admittance matrix formation, matrix factorization, forward and backward substitution, to state visualization, large number of database operations are called repeatedly on data reading, writing, searching, and concurrent accessing. Relational database uses jointly intensive queries for the whole database for many of database operations demanding long computation time for large dataset. On the contrary, graph database outperforms relational database on these database operations [16]. The database operation time on graph dataset is proportional to the number of sub-graphs other than the entire graph leveraging graph database's nodal parallel and hierarchical parallel capabilities.

In this paper, graph database and graph computing technologies will be discussed on their advantages for EMS power flow application. The rest of the paper is organized as follows. The features of graph database are described in Section II. Power flow calculation model and the parallel sparse solver is presented in Section III. Case studies with results are examined in Section IV. Conclusions are provided in Section V.

This work was supported by State Grid Corporation technology project SGRIJSKJ (2016)800

## II. Graph Database Features

Conventional relational database organizes data into tables. Relational database management system (RDBMS) uses Structured Query Language (SQL) for querying and maintaining the database. The database structure stores structured records and their attributes in equal length table. Ideally, data relationships of arbitrary complexity can be presented by relational database. However, the limitations of RDBMS on power system applications are obvious when records have different length of attributes, for example, transmission line and transformer are all branches but have different number and meaning of attributes. In power system applications, they are usually organized into two different tables by relational database.

The relationships between different tables are logical connected by separated table or using joint operation to search common attributes in different tables to find the relationships. For example, creating connectivity of buses and branches requires a join query on bus tables and branch tables or relays on pre-searched and pre-defined connectivity tables.

Maintaining large dataset in RDBMS is challenging. Same data in RDBMS may reside in multiple tables. They are linked to each other through shared key values to represent the data relationships. This design requires multiple operations to add or remove a row (representing add or remove a device in physical grid) in RDBMS. All tables with related shared key need to be updated. Database is evolving when new data and new relationships are added to and/or old data and old relationships are removed from the database. RDBMS does not force table coherence. The flexibility makes engineers easy to define database structure but leaves a hole to create unnecessary complexity on evolving relational databases data growth to analyze larger scale power system with higher fidelity. To deal with complex data structure, queries require sophisticated join operations inviting more computation time.

To accommodate on-line calculation for large scale electric power system, graph database management system (GDBMS) is adapted to fulfill the power system calculation requirements on complex database store, traversal, concurrent access, flexible expansion and reduction. In contrary to relational database, graph database uses graph structures for semantic queries with nodes and edges to store data [17-18]. Unstructured attributes of node or edge are stored in the node or edge. The key concept and merit of GDBMS is the edge directly defined data relationship. The relationships allow stored data to be linked together directly, and be retrieved with one operation other than join operations. This contrasts with RDBMS that manages data in structured tables and linked tables to each other through shared key values. GDBMS, by design, allows simple and fast retrieval of complex hierarchical structures that are difficult to model in RDBMS.

Electric power system is naturally a graph structure. Buses are physically connected through branches as edges just like a graph. The unconstructed parameters of bus, generator, load, branch are stored in node or edge. The graph structure itself naturally represents the topology of electric power grid.

Outperformed than RDBMS, GDBMS supports nodal parallel and hierarchical parallel computation.

### A. Nodal Parallel Computing

In graph computing, nodal parallel computing means computation for each node is independent from each other. They can be performed simultaneously. Figure 1 depicts a power grid connectivity by matrix structure and graph structure. Each numbered vertex represents a bus, the edge between two vertices is a branch with admittance. To format a row of admittance matrix for any vertex, raph-based approach needs to know its neighboring vertex (or vertices) and the edges between them. Each row (corresponding to a node) of admittance matrix can be formed simultaneously.

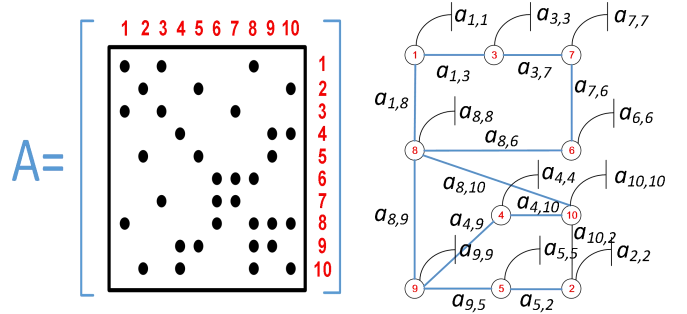

Figure 1. Matrix Structure and Graph Structure for a 10 Bus System

Other examples of nodal parallel computing in power flow calculation are node active power and reactive power injection calculation, node variables mismatch calculation, convergence check at each iteration and post-convergence branch active power and reactive power flow calculation. These calculations on each node are independent to other nodes. This category of parallel computing is defined as nodal parallel computing.

### B. Hierarchical Parallel Computing

Hierarchical parallel computing performs computation on nodes at the same level in parallel. The level next to it is performed after. The hierarchical parallel computing can be applied to matrix factorization, forward and backward substitution. To factorize matrix using Cholesky elimination algorithm, three steps are involved for hierarchical parallel computing: 1) determining fill-ins, 2) forming elimination tree, and 3) partitioning elimination tree for hierarchical parallel computing.

#### 1) Determining Fill-ins

Taking 10 bus system depicting in Figure 1 as an example, the matrix A structure can be represented by graph $G(A)$ below.

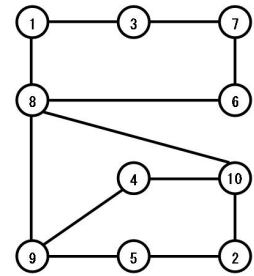

Figure 2. Graph Structure $G(A)$ for Matrix A

The matrix fill-ins during the Cholesky elimination can be determined by the pseudo code of following:

```
FormFilledGraph(Graph A)
{
    for each node i=1 to n of G(A)
        select neighbor node pair j and k of i  //in nodal parallel
            if ( MatrixElement(j,k)=0 and j>i and k>i)
                MatrixElement(j,k)=1;  //fill-in
}
```

MatrixElement(j,k)=1 means element ajk in matrix A is non-zero. The filled graph structure $G^+(A)$ of matrix A with fill-ins after Cholesky elimination is shown in Figure 3. The red edges are fill-ins.

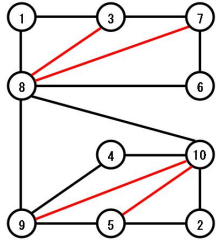

Figure 3.  Filled Graph Structure $G^+(A)$

*2) Forming Elimination Tree*

To parallelize matrix factorization, matrix column dependence needs to be figured out. Elimination tree provides the minimal amount of information on column dependencies in the Cholesky elimination. The pseudo code to form the elimination tree of filled graph structure $G^+(A)$ in parallel by graph computing is shown below and the formed elimination tree of the 10-bus system is shown in Figure 4.

```
FormEliminationTree(FilledGraph A)
{
    Select all nodes i of G+(A)
        Select all neighbor nodes j of i
            ParentOfNode[i] = min(j>i)   //in nodal parallel
}
```

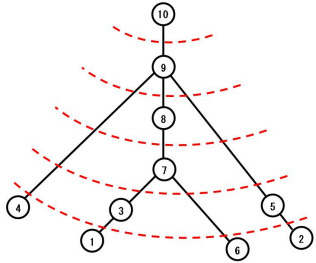

Figure 4.  Elimination Tree T(A)

The pseudo code to create hierarchical partition in parallel by graph computing is shown as follows. And the node hierarchical partition is shown in the Table 1.

```
Level = 1;
FormHierarchicalPartition(EliminationTree A)
{
    Select   all nodes i of T(A) //in nodal parallel
        if ( IsNodeALeave[i] )
            HierarchicalLevel[i] = Level;
            RemoveLeave (i); //remove node i from T(A)
    Level++;
    if (!NoMoreNode)
        FormHierarchicalPartition(Elimination Tree A);
}
```

TABLE I.    NODE PARTITION FOR HIERARCHICAL PARALLEL

| Hierarchical Partition Level | Nodes |
|---|---|
| 1 | 1,2,4,6 |
| 2 | 3,5 |
| 3 | 7 |
| 4 | 8 |
| 5 | 9 |
| 6 | 10 |

During the Cholesky elimination process, forward and backward substitution, the elimination of column i only depends on its parent node on the elimination tree. So the node elimination can be parallelized hierarchically, e.g nodes 1,2,4,6 at the first partition level are processed simultaneously first, the nodes 3,5 at the second partition level secondly till all partitions are processed. The nodal and hierarchical parallel computing is applied to the sparse linear system solver, which will be discussed in the next section

### III. PARALLEL SPARSE SOLVER

Among the applications in EMS, power flow calculation is the fundamental and critical application. There are several different methods to solve the power flow nonlinear equations. The well-known Newton–Raphson method linearizes nonlinear equations using Taylor Series expansion. Industry grade EMS also uses fast decoupled power flow method to approximate active and reactive flow equations to decouple voltage magnitude and angle calculations. Although decoupled power flow method takes a few more iterations than Newton–Raphson method to converge because of the approximated and simplified Jacobian matrix, but each iteration takes much less time. For reactance dominated transmission network, fast decoupled power flow method outperforms Newton–Raphson method on computation efficiency. Usually, in industrial grade EMS, fast decoupled power flow method is conducted first, then Newton–Raphson method if decoupled power flow method is not converged. This strategy practically provides supporting evidence of its effectiveness for contingency analysis for a large-scale system with thousands of contingencies.

In either Newton–Raphson method or fast decoupled power flow method, the power flow problem can be formulated as a sparse linear equation set Ax=b. Forming and solving the equation set consumes about half of power flow computation time. Unlike the tasks of database reading and writing, it is complicated to develop parallel code for A matrix and b vector formation and the solving process of the linear system when using conventional relational database structure. Taking advantage of the nodal and hierarchical parallel computing capability of graph database, parallel computing on forming and solving the sparse linear equation set can be fully exploited. In this section, the parallel processing of A matrix formation, factorization, and forward/backward substitution by graph computing are discussed.

#### A. A Matrix Formation

Using graph database, power system is modeled by graph and defined as an ordered pair of sets (V,E) such that E = {(i,j)|i ∈ V,j ∈ V}. V is vertex set and E is edge set.

Branch and node properties, such as admittance aij and shunt capacitance bii are stored in edges and vertices. Using graph computing, straightforwardly, matrix A can be formed in parallel by the following pseudo-code.

```
FormMatrixA(EdgeSet E, VertexSet V)
{
    Select all nodes i in VertexSet V // in nodal parallel
        Search neighbor nodes j in VertexSet V
            a_ii += a_ij;
}
```

### B. A Matrix Factorization

Matrix A can be factorized as $A = LDU$, where $L$ is the lower triangular matrix, $U$ is the upper triangular matrix, and $D$ is the diagonal matrix. Since matrix A is symmetrical, we get

$$L = U^T \quad (1)$$

$$u_{ij} = l_{ji}, \quad (u_{ij} \in U), \quad (l_{ji} \in L) \quad (2)$$

To solve equation $Ax=b$, factorized matrix is applied as:

$$Ax = LDUx = LDL^T x = b \quad (3)$$

Assume a graph $G(A)$ represents matrix $A$ in graph, where the diagonal element $a_{ii}$ of matrix $A$ represents the attribute of node $i$, and the non-diagonal element $a_{ij} (i \neq j)$ represents the attribute of edge from node $i$ to node $j$. As defined in this section, $G(A)$ is a weighted UDG (Un-Directed Graph). And the factorized graph $G(A')$ of $A$ is a DG (Directed Graph) where two opposite edge $a'_{ij}$ and $a'_{ji}$ in between a pair of nodes are not equal, $a'_{ij} \neq a'_{ji}$ ($i \neq j$).

To form factorized graph $G(A')$, the following pseudo-code is applied for hierarchical parallel computing.

```
FilledGraphA = FormFilledGraph(Graph A);
EliminationTreeA = FormEliminationTree(FilledGraph A);
PartitionP = FormHierarchicalPartition(EliminationTree A);
FactorizeGraphA(EliminationTreeA,A)
{
    for each partition l = 1 to p in Partition P
        Select all nodes i in l //in hierarchical parallel
            a_ij^(n) = a_ij^(n-1) / a_ii^(n-1) ; //in nodal parallel
            a_kj^(n) = a_kj^(n-1) − a_ki^(n-1) · a_ij^(n); //in nodal parallel
}
```

When element in the $i$-th column is normalized, only the nonzero elements are normalized. The eliminations on $i$-node involve all elements on the intersections of the nonzero elements on the $i$-th column and the nonzero elements on the $i$-th row whether the elements at the intersection is non-zero or zero.

The elimination operations on factorized matrix are equivalent to assignment operation on the filled graph $G^+(A)$.

### C. Forward and Backward Substitution by Graph

#### 1) Forward Substitution

Since matrix A is symmetrical, the element $u_{ij}$ in the upper triangular matrix equates to its counterpart $l_{ij}$ in the lower triangular. The code for forward substitution is shown as the follow.

```
ForwardSubstitution(EliminationTree A,L)
{
    for each partition l = 1 to p in Partition P
        Select all nodes i in l //in hierarchical parallel
            Select all neighbor nodes j of i//in nodal parallel
                z_i = b_i − l_ij · z_i; //in nodal parallel
}
```

$z_i$ in the forward substitution is a node attribute, while $l_{ij}$ is an edge attribute in the factorized graph $G(A')$. The forward substitution in graph is to update the attribute $z_i$ with edge attribute $l_{ij}$. According to elimination tree, $z_i$ of node $i$ does not impact to all other nodes but only to these nodes on the path of node $i$. Nodes in the same partition level can be calculated simultaneously. The forward substitution is hierarchical parallel.

#### 2) Normalization

Equivalently, normalization is to solve y in $Dy = z$ as

$$y_i = z_i / d_{ii} \quad (4)$$

where $d_{ii}$ is an element of the diagonal matrix $D$, $z_i$ has been solved in the forward substitution.

In factorized graph $G(A')$, the normalization is to divide the forward substitution result $z_i$ by its attribute $d_{ii}$. Normalization can be performed in nodal parallel.

#### 3) Back Substitution

The code for backward substitution is shown as follows.

```
BackwardSubstitution(EliminationTree A,U)
{
    for each partition l = p to 1 in Partition P
        Select all nodes i in l //in hierarchical parallel
            Select all neighbor nodes j of i//in nodal parallel
                x_i = y_i − u_ij · x_i; //in nodal parallel
}
```

$x_i$ in the backward substitution is a node attribute, while $u_{ij}$ is an edge attribute in the factorized graph $G(A')$. The backward substitution in graph is to update the attribute $x_i$ with edge attribute $u_{ij}$. According to elimination tree, $x_i$ of node $i$ does not impact to all other nodes but only to these nodes on the path of node $i$. Nodes in the same partition level can be calculated simultaneously. The backward substitution can be performed in hierarchical parallel.

## IV. NUMERICAL CASE STUDY

To demonstrate the performance of the proposed graph based parallel online computing, IEEE118 bus system, two provincial systems with 1425 buses and 2643 buses were studied for power flow, contingency analysis, and state estimation. A large system of 10790 buses, 12941 branches, 1588 generators was also studied. We also compared the results with MatPower. Please note the 1425 buses and 2643 buses systems model are not available in MatPower, so the simulation results of the two systems are not available.

The test environment is as listed in Table II.

TABLE II.  HARDWARE AND SOFTWARE TESTING ENVIRONMENT

| Software Environment | |
|---|---|
| Operation System | CentOS6.8 |
| Graph Database | TigerGraph v0.8.1 |
| **Hardware Environment** | |
| CPU | 2CPUs×6 Cores×2 Threads@2.10 GHz |
| Memory | 64GB |

Table 3 summarizes the results of power flow calculation by Newton-Raphson and fast decoupled power flow methods in parallel and in series.

TABLE III.  POWER FLOW CALCULATION RESULTS

| Method | Computing Time (ms) | | | |
|---|---|---|---|---|
| | IEEE 118 Bus System | 1425 Bus System | 2643 Bus System | 10790 Bus System |
| Graph Based Newton | 3.91 | 16.92 | 30.51 | 119.22 |
| Graph Based Fast Decoupled | 2.57 | 12.75 | 24.56 | 86.20 |
| MatPower Newton | 145.7 | NA | NA | 1548 |
| MatPowerFast Decoupled | 30 | NA | NA | 871 |

Shown in the Table III, graph based parallel power flow calculation outperforms the serial conventional power flow calculation for both fast decoupled power flow method and Newton-Raphson method. Under the same convergence criteria, the computation time of power flow in parallel is less than one-tenth of conventional sequential program. The computation efficiency improvement is significant.

## V. CONCLUSION

The trends of electric power system requests power flow calculation to be evolved to accommodate larger scale, higher complex, and uncertain power system with a faster than real-time manner. Taking advantages of modern computation technology and advanced database management, parallel computing is promising to achieve a faster than real-time power flow calculation or even look-ahead capability with future situational awareness.

The proposed graph based power flow calculation in this paper exploited the parallel computing capability using graph computing. Nodal and hierarchical parallelism are developed for common sparse equation solving.

The method underpinnings of the graph based parallel computing is deeply engaged into power flow calculation. The study results on large scale systems show the evidence of the high computation efficiency of the proposed method. The approach proposed in this paper can be adaptable to other power system applications, such as state estimation, contingency analysis, and transient stability analysis and would likely result in significant reduction of computation time.

**Renchang Dai** (M'08, SM'12) received his Ph.D. degree in electrical engineering from Tsinghua University, Beijing, China in 2001. He was with Amdocs, GE Global Research Center, Iowa State University and GE Energy from 2002 to 2017. He is now working at GEIRINA, San Jose, CA involved in smart grid and big data applications.

**Jingjin Wu** received her Ph.D. degree in Computer Science from Illinois Institute of Technology in 2013. Currently, she is a software engineer at GEIRINA in San Jose, California. Before joining GEIRINA, she was a lecturer at University of Electronic Science and Technology of China.

**Chen Yuan** (S'13, M'17) received his Ph.D. degree in electrical and electronics engineering from The Ohio State University, Columbus, OH, USA, in 2016. He is a Postdoctoral Researcher at GEIRI North America. His current research interests include the graph computing, energy management systems, control systems, and smart grid.